\documentclass[letterpaper,english,reprint,superscriptaddress,longbibliography]{revtex4-1}
\usepackage{lmodern}

\usepackage[T1]{fontenc}
\usepackage[utf8]{inputenc}
\usepackage{color}
\usepackage{babel}
\usepackage{float}
\usepackage{units}
\usepackage{mathrsfs}
\usepackage{amsmath}
\usepackage{amssymb}
\usepackage{graphicx}
\usepackage[unicode=true,pdfusetitle,
 bookmarks=false,
 breaklinks=false,pdfborder={0 0 0},pdfborderstyle={},backref=false,colorlinks=false]
 {hyperref}

\makeatletter

\newcommand*\LyXThinSpace{\,\hspace{0pt}}
\pdfpageheight\paperheight
\pdfpagewidth\paperwidth

\makeatother

\begin{document}
\title{Artificial coherent states of light by multi-photon interference in
a single-photon stream}
\author{P. Steindl}
\email{steindl@physics.leidenuniv.nl}

\affiliation{Huygens-Kamerlingh Onnes Laboratory, Leiden University, P.O. Box 9504,
2300 RA Leiden, The Netherlands}
\author{H. Snijders}
\affiliation{Huygens-Kamerlingh Onnes Laboratory, Leiden University, P.O. Box 9504,
2300 RA Leiden, The Netherlands}
\author{G. Westra}
\affiliation{Huygens-Kamerlingh Onnes Laboratory, Leiden University, P.O. Box 9504,
2300 RA Leiden, The Netherlands}
\author{E. Hissink}
\affiliation{Huygens-Kamerlingh Onnes Laboratory, Leiden University, P.O. Box 9504,
2300 RA Leiden, The Netherlands}
\author{K. Iakovlev}
\affiliation{Huygens-Kamerlingh Onnes Laboratory, Leiden University, P.O. Box 9504,
2300 RA Leiden, The Netherlands}
\author{S. Polla}
\affiliation{Huygens-Kamerlingh Onnes Laboratory, Leiden University, P.O. Box 9504,
2300 RA Leiden, The Netherlands}
\author{J.A. Frey}
\affiliation{Department of Physics, University of California, Santa Barbara, California
93106, USA}
\author{J. Norman}
\affiliation{Department of Electrical \& Computer Engineering, University of California,
Santa Barbara, California 93106, USA}
\author{A.C. Gossard}
\affiliation{Department of Electrical \& Computer Engineering, University of California,
Santa Barbara, California 93106, USA}
\author{J.E. Bowers}
\affiliation{Department of Electrical \& Computer Engineering, University of California,
Santa Barbara, California 93106, USA}
\author{D. Bouwmeester}
\affiliation{Huygens-Kamerlingh Onnes Laboratory, Leiden University, P.O. Box 9504,
2300 RA Leiden, The Netherlands}
\affiliation{Department of Physics, University of California, Santa Barbara, California
93106, USA}
\author{W. Löffler}
\email{loeffler@physics.leidenuniv.nl}

\affiliation{Huygens-Kamerlingh Onnes Laboratory, Leiden University, P.O. Box 9504,
2300 RA Leiden, The Netherlands}
\begin{abstract}
Coherent optical states consist of a quantum superposition of different
photon number (Fock) states, but because they do not form an orthogonal
basis, no photon number states can be obtained from it by linear optics.
Here we demonstrate the reverse, by manipulating a random continuous
single-photon stream using quantum interference in an optical Sagnac
loop, we create engineered quantum states of light with tunable photon
statistics, including approximate weak coherent states. We demonstrate
this experimentally using a true single-photon stream produced by
a semiconductor quantum dot in an optical microcavity, and show that
we can obtain light with $g^{(2)}(0)\rightarrow1$ in agreement with
our theory, which can only be explained by quantum interference of
at least 3 photons. The produced artificial light states are, however,
much more complex than coherent states, containing quantum entanglement
of photons, making them a resource for multi-photon entanglement.
\end{abstract}
\maketitle
Coherent states of light are considered to be the most classical
form of light, but expressed in photon number (Fock) space, they consist
of a complex superposition of a  number of photon number (Fock) states.
Because coherent states are non-orthogonal, it is not possible with
linear-optical manipulation and superposition of coherent states to
obtain pure photon number (Fock) states. The opposite is possible
in principle, for instance by attenuating high-$N$ photon number
states one could synthesize coherent states. However, high-$N$ Fock
states are not readily available, but recently high-quality sources
of single-photon ($N=1$) states became accessible based on optical
nonlinearities on the single-photon level. In particular, by using
semiconductor quantum dots in optical microcavities \cite{Santori2002},
single-photon sources with high brightness, purity, and photon indistinguishability
were realized \cite{He2013,Somaschi2016,Ding2016,Snijders2018}. Under
loss, in contrast to higher-$N$ Fock states, single-photon streams
never loose their quantum character since single photons cannot be
split, loss reduces only the brightness. Single photons are an important
resource for quantum information applications \cite{Knill2001}. 

In order to synthesize more complex quantum states of light, multiple
identical single-photon streams can be combined using beamsplitters,
where unavoidably quantum interference appears, the well-known Hong-Ou-Mandel
(HOM) effect \cite{Hong1987}. This effect leads to photon bunching
if the incident photons are indistinguishable, therefore enables the
production of higher photon number states but only probabilistically.
HOM interference is also used for characterization of the photon indistinguishability
of single-photon sources \cite{Santori2002}, which is done mostly
in the pulsed regime where detector time resolution is not an issue.
The regime of a continuous but random stream of single photons has
been explored much less in this aspect, HOM interference with continuous
random stream of true single photons has been observed in Refs. \cite{Patel2008}
and \cite{Proux2015}. The HOM effect can also be used to entangle
photons; in combination with single-photon detection and post-selection,
it also can act as a probabilistic CNOT gate \cite{Larque2008,Fattal2004,Knill2001}. 

\begin{figure}[h]
\includegraphics[width=1\columnwidth]{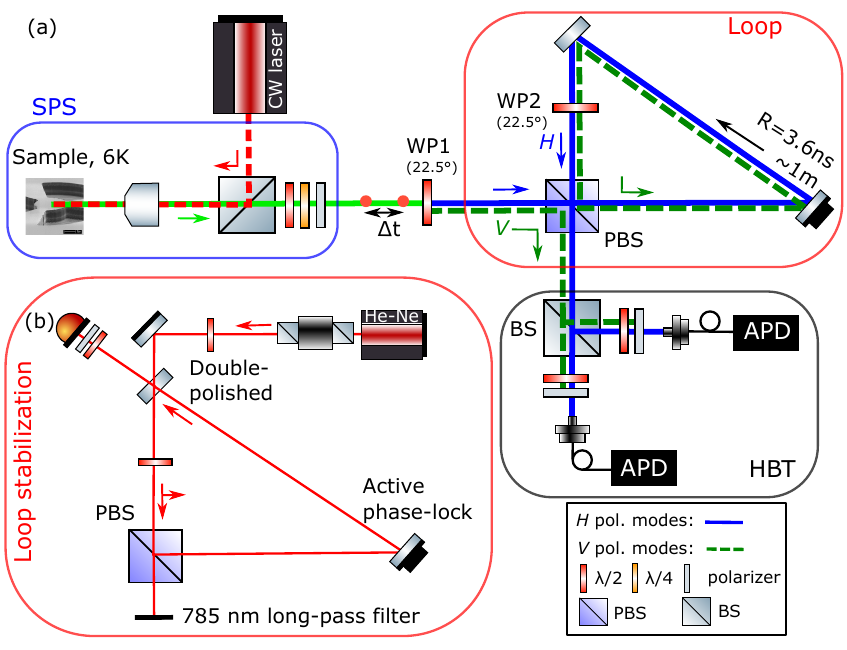}

\caption{Experimental setup - (a) Photons from the single-photon source (SPS)
are diagonally polarized by WP1 before sent to the loop setup consisting
of a polarizing beamsplitter and half-wave plate WP2 at $22.5^{\circ}$.
Light from the loop setup is analyzed with the polarization-resolved
HBT setup. Panel (b) shows the interferometric loop length stabilization.
\label{FIG:Setup}}
\end{figure}

Here we make use of HOM interference in a Sagnac-type delay loop with
a polarizing beamsplitter (Fig. \ref{FIG:Setup}), where HOM interference
happens at a half-wave plate in polarization space \footnote{A half-wave plate with its optical axis at $22.5^{\circ}$ acting
on the two polarization modes is equivalent to the action of a beam
splitter on the two spatial input modes.}. Similar setups are proposed for boson sampling \cite{Motes2014,He2017}
and used for producing linear photonic cluster states \cite{Megidish2012,Pilnyak_PRA17,Istrati2020},
an emerging resource for universal quantum computation \cite{Knill2001,Raussendorf2001,Walther2005}.
Since we operate with a random but continuous single-photon stream,
the repeated quantum interference and enlargement of the spatio-temporal
superposition leads to an infinitely long quantum superposition. By
tuning the photon indistinguishability we observe, in agreement with
our theoretical model, photon correlations approaching that of coherent
light ($g^{(2)}(0)\rightarrow1$), and from our theoretical model,
we deduce that the photon number distribution indeed corresponds to
coherent light, more precisely weak coherent light with a mean photon
number $\bar{n}\approx0.2$. 

Experimentally, as an efficient single-photon source, we use a self-assembled
InGaAs/GaAs quantum dot (QD) embedded in polarization-split micropillar
cavity grown by molecular beam epitaxy \cite{Snijders2018,Snijders2020}.
The QD layer is embedded in a \textit{p-i-n} junction, separated by
a 27\LyXThinSpace nm-thick tunnel barrier from the electron reservoir,
to enable tuning of the QD resonance around 935 nm by the quantum-confined
Stark effect. The QD transition with a cavity-enhanced lifetime of
$\tau_{r}=\unit[130\pm15]{ps}$ is resonantly excited with a continuous-wave
laser, which is separated by a cross-polarization scheme \cite{Snijders2020}
from the single photons that are collected in a single-mode fiber.
This linearly ($H$) polarized single-photon stream $\Psi_{in}$ is
then brought by WP1 ($22.5^{\circ}$) in a superposition of two polarization
modes; $H$-polarized photons enter the 1 m long free-space delay-loop
wherein WP2 ($22.5^{\circ}$) brings them again in a superposition,
only $H$-polarized photons are transmitted from the loop towards
the detection part. Detection is done with a standard Hanbury Brown
and Twiss (HBT) setup with a non-polarizing beamsplitter, after which
the photons are coupled into multi-mode fibers (coupling efficiency
$\sim90$\%) and detected with silicon avalanche photon detectors
(APDs, 25\% efficiency) and analyzed with a time-correlated single-photon
counting computer card. With motorized half-wave plates followed by
a fixed linear polarizer before each multi-mode fiber coupler, the
setup allows to distinguish correlations between photons from the
loop ($g_{HH}^{(2)}(\tau)$), only directly from the source ($g_{VV}^{(2)}(\tau)$),
and to analyze cross-correlations between photons from the loop and
source $g_{VH}^{(2)}(\tau)$. Note that measurement in $VV$ polarization
is equivalent to a standard $g^{(2)}(\tau)$ measurement of the single-photon
source and can be used to obtain a reference without changing the
experimental setup. We have chosen a beam waist of \textbf{$0.50$
}mm inside the loop in order to reduce diffraction loss; the total
round-trip transmission $\eta_{L}$ is $\sim90\%$. Further, we use
active phase-stabilization of the loop length by using a mirror on
a piezoelectric actuator (Fig. \ref{FIG:Setup}(b)) and a frequency-stabilized
He-Ne laser entering the loop through a doubly polished mirror, this
is needed because weak pure single-photon states interfere phase-sensitively
\cite{Loredo2019}. 

\begin{figure}[h]
\includegraphics[width=1\columnwidth]{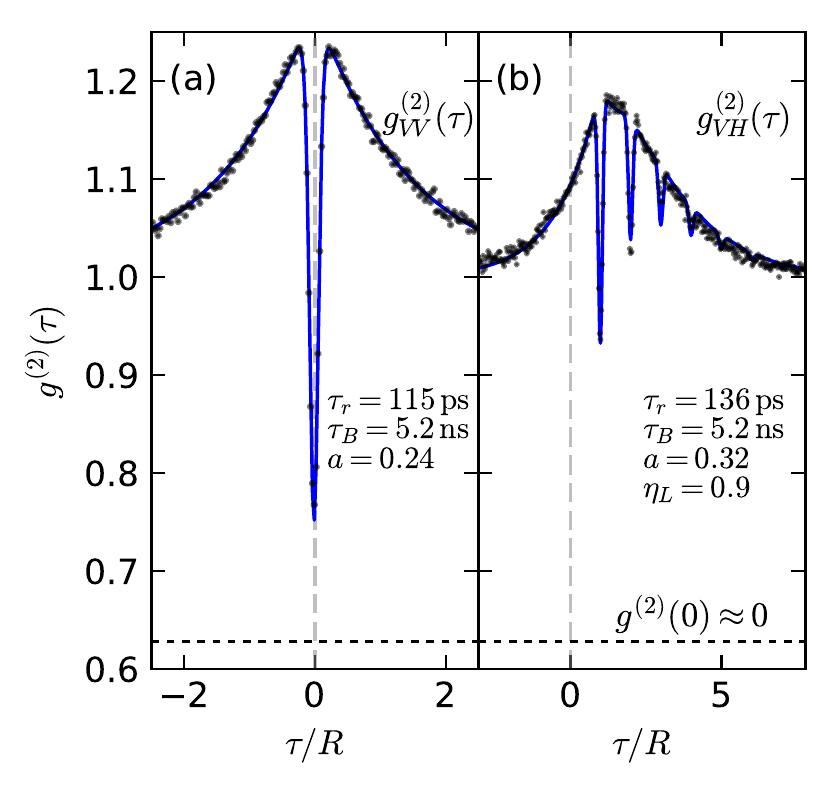}

\caption{Characterization of the single-photon source (a) and loop setup (b),
experimental data has been accumulated over 3 hours; solid lines show
the model calculations. In (a) the three-level model is fitted to
the experimental $g_{VV}^{(2)}(\tau)$ data to obtain the single-photon
source and detector parameters used throughout the paper. Panel (b)
shows $VH$ correlations between photons directly from the source
and from the loop, confirming the validity of our model. \label{FIG:HVandVH}
}
\end{figure}

We operate the QD single-photon source with relatively high excitation
power ($\sim50$ nW) to obtain a bright single-photon stream (detected
single-photon detection rate of $\unit[200]{kHz}$), with the consequence
that unwanted effects produce a broad correlation peak superimposed
to $g^{(2)}(\tau)$. In order to correctly take this into account
in our model, we first measure in $VV$ detector configuration the
source correlations (Fig. \ref{FIG:HVandVH}(a)) and model it using
a three-level system \cite{Kitson_1998,Kurtsiefer_2000}, where $\tau_{B}$
is the lifetime of the additional dark state:
\begin{equation}
g_{3L}^{(2)}(\tau)=1-(1+a)\exp(-|\tau|/\tau_{r})+a\exp(-|\tau|/\tau_{B}).\label{eq:G2_VV}
\end{equation}
Further, for comparison to experimental results with expected $g^{(2)}(0)$
below 0.1 \cite{Snijders2018}, the theoretical data are convolved
with a Gaussian instrument response function (IRF) of our single-photon
detectors with $FWHM=\unit[0.523]{ns}$ \cite{Snijders2016}, limiting
the smallest detectable $g^{(2)}(0)\approx0.63$. From fitting the
model to the experimental data, we obtain a bunching strength $a=0.24\pm0.03$
and $\tau_{B}=\unit[5.2\pm0.3]{ns}$, similar time scales were observed
before \cite{Davanco2014}.

To start building up a theoretical model and to characterize the delay
loop, we now measure in $VH$ detection configuration the cross-correlation
function between photons directly from source and photons from the
delay loop $g_{VH}^{(2)}(\tau)$, shown in Fig. \ref{FIG:HVandVH}(b).
The $V$ detector is connected to the start trigger input of a correlation
card and the $H$ detector to the stop channel, therefore the measured
correlation $g_{VH}^{(2)}(\tau)$ is as expected asymmetric around
$\tau=0$. Considering an $H$-polarized photon entering the loop,
WP2 transforms it into an $\frac{1}{\sqrt{2}}\left(|H\rangle+|V\rangle\right)$
diagonally polarized state. The $H$-polarized part of the state leaves
the loop via the polarizing beamsplitter, while the $V$ part remains
in the loop and is transformed by WP2 into $\frac{1}{\sqrt{2}}\left(|H\rangle-|V\rangle\right)$,
this process is repeating itself infinitely. In the case of a limited
amount of photons in well-defined time bins, the output can easily
be described, the chance that a photon leaves the loop after $r$
round trips is $\left(\nicefrac{\eta_{L}}{2}\right)^{r}$ \cite{Pittman2002}.
In our case of a random single-photon stream, the case is more complex
as we describe the light stream by correlation functions which we
also measure experimentally.
In order to predict $g_{VH}^{(2)}(\tau)$ theoretically, we use as
an approximation that maximally two photons are in the system, which
we prove later to be appropriate here. We obtain for the detected
state for two incident photons with delay $\Delta t\neq0$ (it is
a single-photon source) a weighted superposition of single-photon
streams shifted by time $r\cdot R$, where $r$ is the round-trip
number and $R$ the round-trip delay (see Supplemental Information
\footnote{See Supplemental Material for details of the theoretical derivation of the correlation functions
and further characterization of the artificial coherent states.}):
\begin{equation}
|\Psi_{VH}\rangle=\sum_{\Delta t\neq0}V^{\dagger}\left[\sqrt{\frac{\eta_{L}}{2}}H_{R+\Delta t}^{\dagger}+\sum_{r\geq2}\left(-\sqrt{\frac{\eta_{L}}{2}}\right)^{r}H_{r\cdot R+\Delta t}^{\dagger}\right]|0\rangle.\label{eq:VHstate-maintext}
\end{equation}
The state is written in terms of photon creation operators $V_{t}^{\dagger}$
and $H_{t}^{\dagger}$, where the polarization mode is represented
by the capital letter, the detection time is given in the subscript.
Assuming a source continuously emitting perfect single photons, we
can derive from the two-photon state an analytical expression for
$g_{VH}^{(2)}(\tau)$:
\begin{equation}
g_{VH}^{(2)}(\tau)=1-\sum_{m>0}\left(\frac{\eta_{L}}{2}\right)^{m}\left(1-g_{3L}^{(2)}(\tau-m\cdot R)\right).\label{eq:G2_VH}
\end{equation}

Here, photons with $\Delta t=z\cdot R,z\in\mathbb{Z}$ are correlated
by the loop and create dips in $g_{VH}^{(2)}(\tau)$ for $\tau=m\cdot R$
where $m\in\mathbb{N}$ iterates over round-trips. We observe good
agreement between theory and experimental data in Fig. \ref{FIG:HVandVH}(b).
Note that also the shifted broad peak originating from strong driving
is correctly reproduced.
\begin{figure}[h]
\includegraphics[width=1\columnwidth]{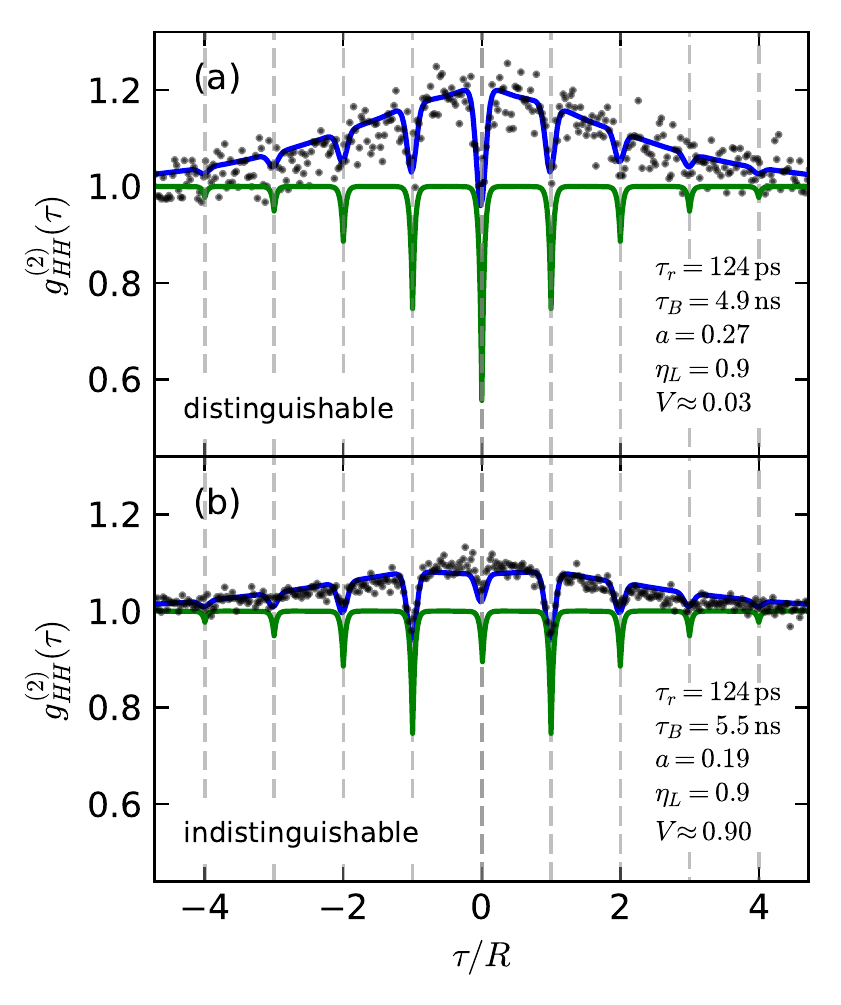}

\caption{Photon correlations $g_{HH}^{(2)}(\tau)$ (symbols) for misaligned
loop (a, $V\approx0.03$, distinguishable photons) and aligned loop
(b, $V\approx0.9$, indistinguishable photons) compared to the model
predictions (blue curves). Raw coincidence counts corresponding to
$g_{HH}^{(2)}(\tau)=1$ were 880 (a) and 9700 (b). The green curves
show the model results for the case without spectral diffusion. \label{FIG:HH}}
\end{figure}

Finally, we investigate the correlations of photons emerging from
the loop by measuring $g_{HH}^{(2)}(\tau)$, shown in Fig. \ref{FIG:HH}.
We find that $g_{HH}^{(2)}(\tau=0)$ is now highly sensitive to the
indistinguishability or wave function overlap $M$ of consecutive
photons produced by the quantum dot, which we can tune experimentally
simply by changing the spatial alignment of the delay loop. Assuming
a perfect single-photon source, the wave function overlap $M$ is
equal to the interferometric visibility $V$, see the Supplemental
section \ref{subsec:Visibility-measurement} for details \cite{Note2}.
The model for the case of distinguishable photons, shown in Fig. \ref{FIG:HH}(a),
can be calculated again in the two-photon picture \cite{Note2}, and
we obtain
\begin{equation}
\begin{aligned}g_{HH}^{(2)}(\tau)= & 1-\frac{2\eta_{L}}{4-\eta_{L}^{2}}\sum_{m\in\mathbb{Z}\backslash\left\{ 0\right\} }\left(\frac{\eta_{L}}{2}\right)^{|m|}\left(1-g_{3L}^{(2)}(\tau-m\cdot R)\right)\\
 & -\left(1-g_{HH}^{(2)}(0)\right)\left(1-g_{3L}^{(2)}(\tau)\right),
\end{aligned}
\label{eq:G2_HH}
\end{equation}
where the value of $g_{HH}^{(2)}(0)$ has to be calculated using full
quantum state propagation which we describe now.

The delay loop leads to quantum interference of photons at different
times in the incident single-photon stream, and HOM photon bunching
occurring at WP2 produces higher photon number states in a complex
quantum superposition. We have developed a computer algorithm that
can simulate $g_{HH}^{(2)}(0)$, see the Supplemental Information
section \ref{subsec:Simulation-details} for details \cite{Note2}.
For the results shown here, we take up to 20 photons or loop iterations
into account to approximate the experiment with a continuous photon
stream.\textbf{ }For completely distinguishable photons we obtain
$g_{HH}^{(2)}(0)=0.49$ (corrected for dark state dynamics), which
agrees well with the experimentally observed correlations in Fig.
\ref{FIG:HH}(a). 

\begin{figure}[h]
\includegraphics[width=1\columnwidth]{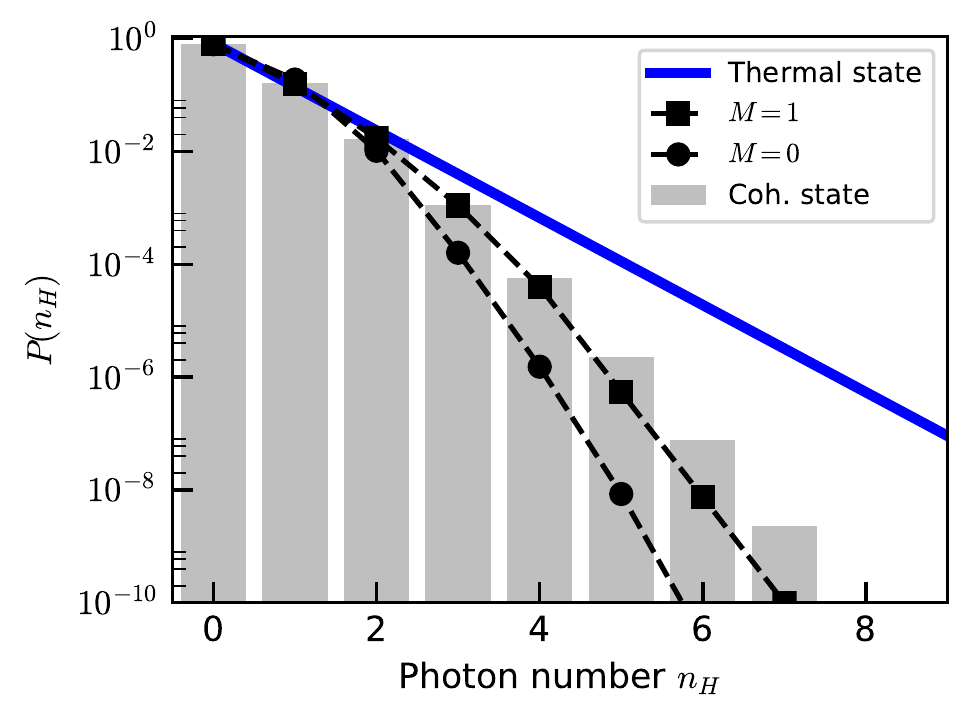}

\caption{Comparison of the photon number distribution of a weak coherent state
(bars) and\textcolor{black}{{} thermal state (through blue line) with
same mean photon number $\overline{n}\approx0.2$, to the results
from our theoretical model (squares for the case of indistinguishable
photons, $M=1$, and circles for the case of distinguishable photons,
$M=0$). A fixed round-trip loss of 0.1 is included in both cases.
The artificial state matches best to the weak coherent light state.}
\label{FIG:HH-prob} }
\end{figure}

\begin{figure}[h]
\includegraphics[width=1\columnwidth]{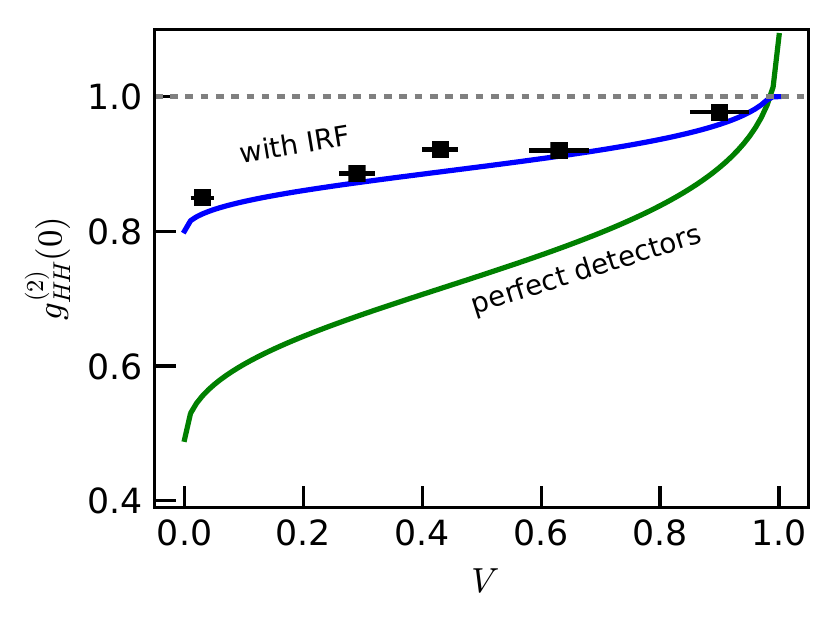}

\caption{Experimental $g_{HH}^{(2)}(0)$ (black squares), corrected for dark
state dynamics, compared to our theoretical expectations (blue line,\textbf{
}convolved with IRF) as a function of the photon indistinguishability
or wave-function overlap, expressed by the visibility $V$.\textbf{
}Light with coherent photon statistics is obtained for $V\rightarrow1$.
The green line shows the expected results for detectors with perfect
timing resolution. Results include a fixed round-trip loss of 0.1.\label{FIG:HHvsM}}
\end{figure}

For the case of indistinguishable photons with maximal wave-function
overlap $M\approx0.9$, we observe in Fig. \ref{FIG:HH}(b) that the
dip at $\tau=0$ almost disappears. This is because the (multi-)photon
bunching increases the weight of higher photon number states, and,
as we show now, produces quasi-coherent states of light with $g^{(2)}(0)\approx1$.
Based on our computer simulation, we investigate the photon number
distribution $P(n)$, which is shown in Fig. \ref{FIG:HH-prob}. We
see very good agreement of the artificial coherent state (indistinguishable
photons, $M=1$, experimentally we achieve $M=0.9$) to an exact weak
coherent light state with the same mean photon number ($\bar{n}=0.2$).
In the Supplemental Information \cite{Note2} Section \ref{subsec:Artificial-coherent-state-properties}
we show that the artificial coherent state is also very close to being
an eigenstate of the annihilation operator, as expected. Now, using
the full simulated quantum state, we calculate the quantum fidelity
$F$ to the exact coherent state and obtain\textbf{ }$1-F\approx10^{-3}$
for both\textbf{ $M=1$} and \textbf{$M=0.9$}. We also calculate
the $l_{1}$-norm of coherence \cite{Baumgratz2014} $C_{l_{1}}$,
also here the deviation from the exact coherent state is very small,
smaller than $10^{-3}$ relatively. From comparison of the density
matrices \cite{Note2}, we see that deviations occur mainly in the
higher photon number components, those are weak and do not contribute
much to the aforementioned measures. These small deviations are also
visible in the Wigner function of the artificial coherent state \cite{Note2}.

In the model, we can ignore a round-trip dependent decrease of $M$
due to beam diffraction since the effect is only $\sim2\unit{\%}$,
see Supplemental section\ref{subsec:Delay-loop:-Round-trip}, and
from Fig. \ref{FIG:HH-prob} we also see why it was justified above
to ignore $N>2$ states for prediction of $g_{VH}^{(2)}(\tau)$ and
$g_{HH}^{(2)}(\tau\neq0)$, their contribution is negligible (Supplemental
section \ref{subsec:How-many-photons} \cite{Note2}). In our experiment,
we can also observe the transition to an artificial coherent state
by tuning the photon indistinguishability $M$ to intermediate values,
which is shown in Fig. \ref{FIG:HHvsM}, again in good agreement with
our model. Compared to a weak thermal state of light which can be
produced by spontaneous emission of many single-photon emitters coupled
to the same cavity mode \cite{Hennrich2005}, although having similar
$P(n)$ for low $\bar{n}$, as shown in Fig. \ref{FIG:HH-prob}, $g^{(2)}(0)$
would show a peak which is not the case here. The simple characterization
method based only on two-photon correlations measurement presented
here could also be useful for characterization of photonic cluster
states demonstrated recently \cite{Pilnyak_PRA17,Istrati2020}. In
order to determine how many photons are contributing to the quasi-coherent
states here, by comparing our experimental results to a photon-truncated
theoretical model, we see that at least 3 photons are needed to explain
our results. We estimate that these three-photon states occur with
a rate of about $5$ kHz in our experiment \cite{Note2}. 

In conclusion, we have shown approximate synthesis of continuous-wave
coherent states of light from a quantum dot-based single-photon source,
using a simple optical setup with a free-space delay loop. The underlying
mechanism is repetitive single-photon addition \cite{Barnett2018,Kim2005,Marco2010}
to an ever-growing number-state superposition, and can be tuned by
changing photon distinguishability. A difference of the artificial
coherent states here to conventional coherent light is that the photons
of the artificial coherent state are correlated with others separated
by multiples of the loop delay, this is typical for systems with time-delayed
feedback \cite{Pototsky2007} including lasers \cite{Holzinger2018,Wang2020}.
This quantum entanglement becomes accessible if an ordered (pulsed)
stream of single photons is used, and enables production of linear
cluster states which has been realized recently \cite{Pilnyak_PRA17,Istrati2020},
and feed-forward or fast modulators \cite{Rohde2015,He2017,Marek2018,Svarc2020}
can be used to produce even more complex quantum states. We want
to add that also lasers produce only approximately coherent states
with entanglement of the stimulated photons via the gain medium \cite{Molmer1997,Enk2002,Gea-Banacloche2002,Pegg2005}
which is in practice inaccessible due to the impossibility of monitoring
every quantum interaction in the system \cite{Noh2008}. From this
quantum entanglement arises complexity, therefore we had to use algorithmic
modelling in order to produce a theoretical prediction of the output
state; this is not surprising because it is known to be computationally
hard to calculate quantum interference with many beamsplitters (including
loop setups such as the one investigated here) and many photons in
Fock states, possibly lying beyond the $P$ complexity class \cite{Gard2014,Motes2014}.
It would be an interesting goal to develop rigorous entanglement (length)
witnesses that can also be applied to continuous and random photon
streams such as here, explore possibilities for time-bin encoded tensor
networks \cite{Lubasch2018,Dhand2018} or quantum metrology \cite{Peniakov2020},
or to entangle the photons in a $d>1$-dimensional topology \cite{Asavanant2019,Larsen2019}.
A natural question is if other quantum states of light, in particular
quadrature squeezed light, can be produced in a similar way, unfortunately,
those light states are not resilient against loss compared to coherent
states, rendering this far more challenging. 
\begin{acknowledgments}
We thank Gerard Nienhuis, Rene Allerstorfer and Harmen van der Meer
for discussions and support, and we acknowledge funding from the European
Union’s Horizon 2020 research and innovation programme under grant
agreement No. 862035 (QLUSTER), from FOM-NWO (08QIP6-2), from NWO/OCW
as part of the Frontiers of Nanoscience program and the Quantum Software
Consortium, and from the National Science Foundation (NSF) (0901886,
0960331).
\end{acknowledgments}

%\bibliographystyle{../naturemagwV1allauthors}
%\bibliography{../cw_loop_bib}

\pagebreak \onecolumngrid\renewcommand{\thefigure}{S\arabic{figure}}\setcounter{figure}{0}\renewcommand{\theequation}{S\arabic{equation}}\setcounter{equation}{0}

\section*{Supplemental Information}

\subsection{The two-photon picture\label{subsec:Two-photon-picture}}

In this section, we derive expressions for $g_{VH}^{(2)}(\tau)$ and
$g_{HH}^{(2)}(\tau)$ presented in the main text. Limiting the description
to two distinguishable photons (see main text), we first describe
the full two-photon state created from the single-photon stream by
the delay loop. Later, we derive the polarization-postselected correlation
functions. Finally, we describe how we include the finite lifetime
of the single-photon source, as well as imperfections such as quantum
dot blinking and photon loss in the delay loop. Later, in section
\ref{subsec:Simulation-details} we will discuss indistinguishable
photons and the effect of photon bunching by the Hong-Ou-Mandel effect.

A single photon entering the loop setup (with the round-trip delay
$R$) shown in Fig. \ref{FIG:Setup} at time $t$ is brought into
a quantum superposition which becomes increasingly complex with the
number of round-trips $r$. The single-photon state $|\Psi_{1}\rangle$
can be written in terms of photon creation operators $Q^{\dagger}$
acting on the vacuum as $|\Psi_{1}\rangle=Q^{\dagger}(r,t)|0\rangle,$
where $Q^{\dagger}(r,t)$ depends on the number of round-trips:

\begin{equation}
Q^{\dagger}(r,t)=\begin{cases}
\frac{{1}}{\sqrt{{2}}}V_{D,t}^{\dagger}+\frac{{1}}{\sqrt{{2}}}H_{B,t}^{\dagger}, & \mathrm{for}\:r=0,\\
\frac{{1}}{\sqrt{{2}}}V_{D,t}^{\dagger}+\frac{{1}}{{2}}\left(H_{D,t+R}^{\dagger}+V_{B,t+R}^{\dagger}\right), & r=1,\\
\frac{{1}}{\sqrt{{2}}}V_{D,t}^{\dagger}+\frac{{1}}{{2}}H_{D,t+R}^{\dagger}+\frac{1}{\sqrt{2}}\left[\sum_{s\geq2}^{r}\left(\frac{-1}{\sqrt{2}}\right)^{s}H_{D,t+s\cdot R}^{\dagger}-\left(\frac{-1}{\sqrt{2}}\right)^{r}V_{B,t+r\cdot R}^{\dagger}\right], & r\geq2.
\end{cases}\label{eq:LoopCreatingOper}
\end{equation}
Here, for instance, $V_{D,t}^{\dagger}$ is the photon creation operator
for a $V$-polarized photon in mode $D$ at time $t$. In each round
trip, a photon at position $B$ is brought into quantum superposition
by WP2, and the $H$-polarized component is transmitted from the loop
by the polarizing beamsplitter (PBS) to the HBT detection setup, this
results in an infinite tree-like structure indicated in Fig.\ref{FIG:Algorithm_block_chart}
below.

We assume that the input light is a perfect but random single-photon
stream, where the delay between two photons is $\Delta t\neq0$, and
the HBT setup post-selects two-photon detection events from this stream.
In general, this (detected) two-photon state can be written as
\begin{equation}
|\Psi_{2}\rangle=\sum_{\Delta t\neq0}Q^{\dagger}(r_{1},t)\otimes Q^{\dagger}(r_{2},t+\Delta t)|0\rangle.\label{eq:two-photon-state}
\end{equation}
Since only photons in spatial mode $D$ are detected, we ignore other
photons and leave out the spatial label from now on. In this section,
we assume a perfect single-photon source and no loss in the delay
loop.

\subsubsection{Detection of $VH$ correlations}

Working out Eq. (\ref{eq:two-photon-state}) explicitly and post-selecting
on terms containing one $V$ and one $H$ photon, we obtain

\begin{equation}
|\Psi_{VH}\rangle=\frac{1}{2}\sum_{\Delta t\neq0}\left[\frac{1}{\sqrt{2}}V_{t}^{\dagger}H_{t+\Delta t+R}^{\dagger}+V_{t}^{\dagger}\sum_{r_{2}\geq2}\left(\frac{-1}{\sqrt{2}}\right)^{r_{2}}H_{t+\Delta t+r_{2}\cdot R}^{\dagger}+\frac{1}{\sqrt{2}}V_{t+\Delta t}^{\dagger}H_{t+R}^{\dagger}+V_{t+\Delta t}^{\dagger}\sum_{r_{1}\geq2}\left(\frac{-1}{\sqrt{2}}\right)^{r_{1}}H_{t+r_{1}\cdot R}^{\dagger}\right]|0\rangle.\label{eq:VHstate-step0}
\end{equation}
Now, we consider that $V$ photons start the time-correlated single-photon
counting apparatus at time $t=0$, therefore we require that in the
first two terms $t=0$ and in the last two terms $t+\Delta t=0$.
Obviously, only photons with $\Delta t=z\cdot R,z\in\mathbb{Z}$ are
correlated by the loop, the rest is uncorrelated and contributes to
the correlation function as $g_{VH}^{(2)}(\tau\neq z\cdot R)=1$.
Due to symmetry between $r_{1}$ and $r_{2}$, we see that the final
state is just a weighted superposition of single-photon streams shifted
with respect to each other by a time $r_{2}\cdot R,r_{2}\in\mathcal{\mathscr{\mathtt{\mathbb{N}}}}$:
\begin{equation}
|\Psi_{VH}\rangle=\sum_{\Delta t\neq0}V^{\dagger}\left[\frac{1}{\sqrt{2}}H_{R+\Delta t}^{\dagger}+\sum_{r_{2}\geq2}\left(\frac{-1}{\sqrt{2}}\right)^{r_{2}}H_{r_{2}\cdot R+\Delta t}^{\dagger}\right]|0\rangle.\label{eq:VHstate}
\end{equation}
Evaluating the state for individual data points of $g_{VH}^{(2)}(\tau)$
for fixed $\tau=\mu\cdot R$, we obtain for each $\mu$ an analytical
expression (normalized by $g_{VH}^{(2)}(\pm\infty)=1$), depending
on whether the photons are correlated by the loop:
\begin{equation}
g_{VH}^{(2)}(\tau=\mu\cdot R)=\begin{cases}
\sum_{z=1}2^{-z}-2^{-\mu}=1-2^{-\mu} & \mathrm{for\:}\mu=m\in\mathbb{N},\\
\sum_{z=1}2^{-z}=1 & \mathrm{else}.
\end{cases}\label{eq:gvh-individualm}
\end{equation}
By summation over time we obtain the full form for $g_{VH}^{(2)}(\tau)$
\begin{equation}
g_{VH}^{(2)}(\tau)=\sum_{\mu}g_{VH}^{(2)}(\tau=\mu\cdot R)\delta(\tau-\mu\cdot R)=1-\sum_{m\in\mathbb{N}}2^{-m}\delta(\tau-m\cdot R).\label{eq:gvh-lossfree}
\end{equation}

\subsubsection{Detection of $HH$correlations \label{subsec:correlationsHH}}

If both detectors detect $H$-polarized photons, all detected photons
must have come from the loop. We can post-select $H$-polarized photons
from Eq. (\ref{eq:two-photon-state}) and write the two-photon (not
normalized) state using $t'=t+R$ as

\begin{equation}
|\Psi_{HH}\rangle=\frac{1}{2}\sum_{\Delta t\neq0}\sum_{r_{1}\geq1}\sum_{r_{2}\geq1}\left(\frac{-1}{\sqrt{2}}\right)^{r_{1}+r_{2}}H_{t'+(r_{1}-1)\cdot R}^{\dagger}H_{t'+\Delta t+R(r_{2}-1)}^{\dagger}|0\rangle.\label{eq:HHstate}
\end{equation}
We again consider that photons from the single-photon stream separated
by $\Delta t=z\cdot R$ are correlated by the loop, and coincidence
clicks are recorded with time delay $\tau=m\cdot R,m\in\mathbb{Z}$.
This leads to the condition $z=r_{1}+r_{2}\pm m$. For normalization
of the second-order correlation function, we require the coincidence
probability for photon delays different than the loop delay. We define
$0<\epsilon<1$ and $z'=z+\epsilon$ and $m'=m+\epsilon$ and assume
the conditions as before. We start the calculation of $g^{(2)}(\tau)$
by considering correlations at $\tau=\mu\cdot R$ for loop-correlated
photons ($\mu=m$), which results in

\begin{equation}
g_{\mathrm{cor}}^{(2)}(\tau=\mu\cdot R)=\sum_{z\neq0}\sum_{r_{1}\geq1}\sum_{r_{2}\geq1}\frac{\langle\Psi_{HH}(r_{1},r_{2},\Delta t)|\Psi_{HH}\rangle}{\langle\Psi_{HH}|\Psi_{HH}\rangle}\cdot\delta\left(z-r_{1}+r_{2}\pm\mu\right).\label{eq:ghh-cor}
\end{equation}
The state $|\Psi_{HH}(r_{1},r_{2},\Delta t)\rangle$ is a superposition
of two photons with fixed roundtrips $r_{1}$ and $r_{2}$ and fixed
$\Delta t$ (summand in Eq. (\ref{eq:HHstate})). Similarly, uncorrelated
photons contribute to the correlations only if $\mu=m'\neq m$ (i.e.
$\epsilon\neq0$):
\begin{equation}
g_{\mathrm{uncor}}^{(2)}\left(\tau=\mu\cdot R\right)=\sum_{z\in\mathbb{Z}}\sum_{r_{1}\geq1}\sum_{r_{2}\geq1}\frac{\langle\Psi_{HH}\left(r_{1},r_{2},\Delta t\right)|\Psi_{HH}\rangle}{\langle\Psi_{HH}|\Psi_{HH}\rangle}\cdot\delta\left(z-r_{1}+r_{2}\pm\mu\right).\label{eq:ghh-uncor}
\end{equation}
The only difference between the equations above is in the summation
over $z$. Since we deal with a single-photon source which implies
$\Delta t\neq0$, the sum in Eq. (\ref{eq:ghh-cor}) must not include
$z=0$, while this is naturally satisfied in Eq. (\ref{eq:ghh-uncor})
where we can sum over all integer numbers $\mathbb{Z}$. 

Polarization post-selection is a non-unitary operation, therefore
the state and also the resulting correlations are not normalized.
Hence, we follow the usual normalization procedure, i.e., we normalize
by the uncorrelated correlations $g_{\mathrm{uncor}}^{(2)}\left(\tau=\mu\cdot R\right)=1$.
Moreover, this choice also helps to simplify the infinite series,
where the double summation over $r_{1}$ and $r_{2}$ can be evaluated
and we obtain for fixed $\mu$ 
\begin{equation}
\begin{aligned}g_{HH}^{(2)}(\tau=\mu\cdot R)=\begin{cases}
\frac{g_{\mathrm{cor}}^{(2)}(\tau=\mu\cdot R)}{g_{\mathrm{uncor}}^{(2)}\left(\tau=\mu\cdot R\right)}=1-2\left(\frac{1}{2}\right)^{|m|}\sum_{r_{1\geq1}}2^{-2r_{1}}=1-\frac{2}{3}\left(\frac{1}{2}\right)^{|m|}, & \mathrm{for}\:\mu=m\\
\frac{g_{\mathrm{uncor}}^{(2)}\left(\tau=\mu\cdot R\right)}{g_{\mathrm{uncor}}^{(2)}\left(\tau=\mu\cdot R\right)}=1, & \mathrm{for}\:\mu\neq m,
\end{cases}\end{aligned}
\label{eq:ghh-individualm}
\end{equation}
where the sum gives a factor of $\nicefrac{1}{3}$. We then obtain
the full, loop-loss free ideal correlation function 
\begin{equation}
g_{HH}^{(2)}(\tau)=\sum_{\mu}g_{HH}^{(2)}(\tau=\mu\cdot R)\delta(\tau-\mu\cdot R)=1-\frac{2}{3}\sum_{m\in\mathbb{Z}}\left(\frac{1}{2}\right)^{|m|}\delta(\tau-m\cdot R).\label{eq:ghh-lossfree}
\end{equation}

\subsubsection{Relation to source correlations}

As written in the main text, we take the finite quantum dot lifetime
and blinking into account by the single-photon source correlation
function $g_{3L}^{(2)}(\tau)$ \cite{Kurtsiefer_2000_SM} in Eq. (\ref{eq:G2_VV}).
In order to include this in the model, we replace the $\delta$-function
in Eqs. (\ref{eq:gvh-lossfree}) and (\ref{eq:ghh-lossfree}) by the
source correlation function like

\[
\delta(\tau-m\cdot R)\rightarrow\exp\left(-\frac{|\tau-m\cdot R|}{\tau_{r}}\right)\rightarrow\left(1-g_{3L}^{(2)}(\tau-m\cdot R)\right).
\]

The first replacement would include only the finite lifetime, while
the second includes also blinking. We obtain (note that we here and
in the following re-define $g_{VH}^{(2)}(\tau)$ and $g_{HH}^{(2)}(\tau)$
as we develop the model):

\begin{equation}
g_{VH}^{(2)}(\tau)=1-\frac{1}{2}\sum_{m>0}\left(\frac{1}{2}\right)^{m}\left(1-g_{3L}^{(2)}(\tau-m\cdot R)\right)\label{eq:gvh-lossfree-lifetime}
\end{equation}
and
\begin{equation}
g_{HH}^{(2)}(\tau)=1-\frac{2}{3}\sum_{m\in\mathbb{Z}}\left(\frac{1}{2}\right)^{|m|}\left(1-g_{3L}^{(2)}(\tau-m\cdot R)\right).\label{eq:ghh-lossfree-lifetime}
\end{equation}

\subsubsection{Loss in the delay loop \label{subsec:correlationsLOSS}}

Understanding the effects of optical loss in the delay loop is essential
for correct modelling of the produced quantum state of light. In order
to achieve this, we define the loop transmission as $\eta_{L}$ and
incorporate it in the two-photon state of Eq. (\ref{eq:VHstate})
for each round trip simply by replacing $\sqrt{1/2}$ by $\sqrt{\eta_{L}/2}$:

\begin{equation}
|\Psi_{VH}\rangle=\sum_{\Delta t\neq0}V^{\dagger}\left[\sqrt{\eta_{L}/2}H_{R+\Delta t}^{\dagger}+\sum_{r_{2}\geq2}\left(-\sqrt{\eta_{L}/2}\right)^{r_{2}}H_{r_{2}\cdot R+\Delta t}^{\dagger}\right]|0\rangle.\label{eq:VHstateLoss}
\end{equation}

After this, the correlation function $g_{VH}^{(2)}(\tau=\mu\cdot R)$
in Eq. (\ref{eq:gvh-individualm}) has to be re-normalized for each
$\mu$
\begin{equation}
g_{VH}^{(2)}(\tau=\mu\cdot R)=\begin{cases}
1-\left(\frac{\eta_{L}}{2}\right)^{\mu}=1-\left(\frac{\eta_{L}}{2}\right)^{\mu} & \mathrm{for\:}\mu=m\in\mathbb{N},\\
1 & \mathrm{else}.
\end{cases}\label{eq:gvh-individualm-wloss}
\end{equation}
Similar as before, we obtain the full $g_{VH}^{(2)}(\tau)$ by adding
it up for all $\mu$, and, after inserting $g_{3L}^{(2)}(\tau)$ we
obtain
\begin{equation}
g_{VH}^{(2)}(\tau)=1-\sum_{m>0}\left(\frac{\eta_{L}}{2}\right)^{m}\left(1-g_{3L}^{(2)}(\tau-m\cdot R)\right).\label{eq:gvh-final}
\end{equation}
Analogously, in order to include loop loss in $g_{HH}^{(2)}(\tau)$,
we first replace in $|\Psi_{HH}\rangle$ (Eq. (\ref{eq:HHstate}))
$\sqrt{1/2}$ by $\sqrt{\eta_{L}/2}$. This change equally affects
$g_{\mathrm{cor}}^{(2)}\left(\tau=\mu\cdot R\right)$ and $g_{\mathrm{uncor}}^{(2)}\left(\tau=\mu\cdot R\right)$,
allowing us to again normalize by $g_{\mathrm{uncor}}^{(2)}\left(\tau=\mu\cdot R\right)$.
In analogy to Eq. (\ref{eq:ghh-individualm}) we obtain

\begin{equation}
\begin{aligned}g_{HH}^{(2)}(\tau=\mu\cdot R)=\begin{cases}
\frac{g_{\mathrm{cor}}^{(2)}(\tau=\mu\cdot R)}{g_{\mathrm{uncor}}^{(2)}\left(\tau=\mu\cdot R\right)}=1-2\left(\frac{\eta_{L}}{2}\right)^{|m|}\sum_{r_{1\geq1}}\left(\frac{\eta_{L}}{2}\right)^{2r_{1}}=1-\frac{2\eta_{L}}{4-\eta_{L}^{2}}\left(\frac{\eta_{L}}{2}\right)^{|m|}, & \mathrm{for}\:\mu=m\\
\frac{g_{\mathrm{uncor}}^{(2)}\left(\tau=\mu\cdot R\right)}{g_{\mathrm{uncor}}^{(2)}\left(\tau=\mu\cdot R\right)}=1, & \mathrm{for}\:\mu\neq m,
\end{cases}\end{aligned}
\label{eq:ghh-individualm-wloss}
\end{equation}
and finally complete expression for $g_{HH}^{(2)}(\tau)$ including
loop loss:
\begin{equation}
g_{HH}^{(2)}(\tau)=1-\frac{2\eta_{L}}{4-\eta_{L}^{2}}\sum_{m\in\mathbb{Z}}\left(\frac{\eta_{L}}{2}\right)^{|m|}\left(1-g_{3L}^{(2)}(\tau-m\cdot R)\right).\label{eq:ghh-final}
\end{equation}

Finally, based on the experimental observation that $g_{HH}^{(2)}(\tau)$
is only for $\tau=0$ sensitive to multi-photon quantum interference,
we explicitly include $g_{HH}^{(2)}(0)$ and arrive at Eq. (\ref{eq:G2_HH})
of the main text. We explain the numeric calculation of $g_{HH}^{(2)}(0)$
in the next section.

\subsection{Simulation details}

\label{subsec:Simulation-details} Calculation of $g_{HH}^{(2)}(0)$
for indistinguishable photons is complex due to multi-photon quantum
interference, we accomplish this by a computer simulation that iteratively
calculates the evolution of $H$-polarized photons in the experiment
in Fig. \ref{FIG:Setup}. Because we aim to simulate and tune quantum
interference at WP2 by misaligning the delay loop, we introduce two
spatial bases for the description of spatially separated photons on
WP2. The first basis $\left\{ H,V\right\} _{S}$ describes the incident
photons and, because $H$-polarized photons are only detected after
at least one round-trip, also after the first round trip; while $\left\{ H,V\right\} _{L}$
is the basis used to represent the state after the second round trip.
Below, the spatial mode of polarized photons is stressed by its subscript. 

\begin{figure}[h]
	\includegraphics[width=0.6\columnwidth]{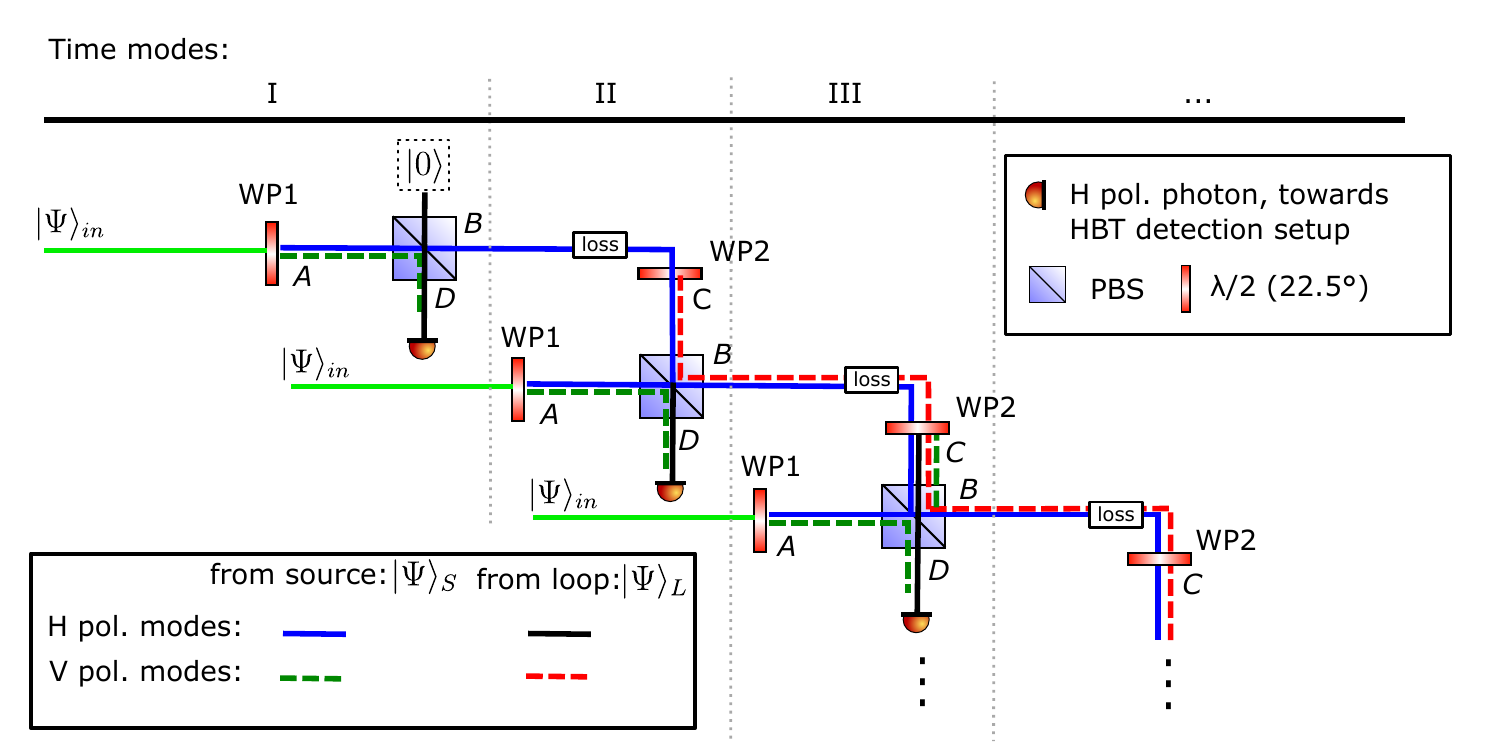}
	
	\caption{Flow chart of the computer algorithm to calculate $g_{HH}^{(2)}(0)$.
		It is an unfolded version of the real experiment where photons in
		mode $D$ are sent to the HBT detection setup, each of these ports
		corresponds to a different time mode. \label{FIG:Algorithm_block_chart}}
\end{figure}

In the simulation, we assume a perfect single-photon source continuously
emitting $H$-polarized photons with mutual delay of $\Delta t=R$,
each photon corresponds to $|\Psi\rangle_{in}$ in the flow chart
of the algorithm in Fig. \ref{FIG:Algorithm_block_chart}. WP1 set
to $22.5^{\circ}$ transforms each $H$-polarized photon to $\frac{1}{\sqrt{2}}(|H_{S}\rangle+|V_{S}\rangle)$,
the $V_{S}$ mode is then erased, while the $H_{S}$ mode is transformed
by the PBS from input \textit{A} to \textit{\emph{output }}\textit{B}
and enters the (initially empty) optical loop. The $H$ mode in the
loop is then transformed on WP2 ($22.5^{\circ}$) into $\frac{1}{\sqrt{2}}(|H_{S}\rangle+|V_{S}\rangle)$,
arrives at time $R$ at port \textit{C} of the PBS\textit{ }and is
transformed from the $\left\{ H,V\right\} _{S}$ to the $\left\{ H,V\right\} _{L}$
basis. At the same time, the next photon from input \textit{A} arrives
at the PBS and its $H$-polarized part is sent to the loop, while
outgoing photons in port\textit{ D} are sent to the HBT detection
setup (Fig. \ref{FIG:Setup} in the main text), where different ports
\textit{D} in Fig. \ref{FIG:Algorithm_block_chart} correspond to
different time bins. 

\begin{figure}[h]
	\includegraphics[width=0.5\columnwidth]{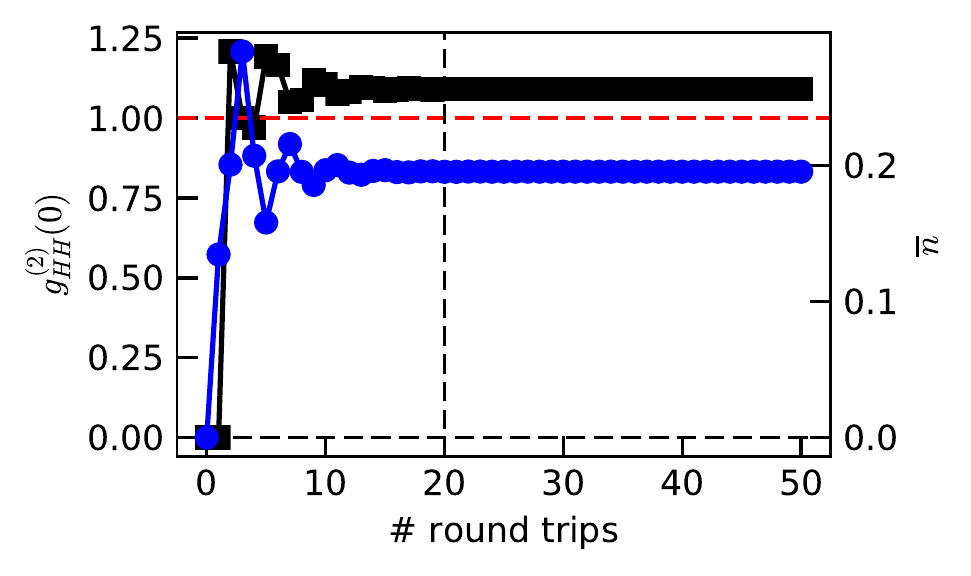}
	
	\caption{Correlation $g_{HH}^{(2)}(0)$ (black) and average photon number $\overline{n}$
		(blue) as a function of the number of round trips for perfectly indistinguishable
		photons ($M=1$) and experimental round-trip loss $\eta_{L}=0.9$.
		The vertical line shows our choice of 20 round trips where the field
		is sufficiently converged.\label{FIG:FieldConvergence}}
\end{figure}

In order to model the continuous time-averaged measurements, we have
to create a stable photonic field in the initially empty delay loop.
In Fig. \ref{FIG:FieldConvergence} we show how the average photon
number $\overline{n}$ and $g_{HH}^{(2)}(0)$ evolve with the number
of round trips. We observe initial fluctuations in both parameters
and very good convergence from 20 round trips on, which we use for
all calculations in this paper.

\subsubsection{Quantum interference at waveplate WP2 in the delay loop }

As explained in the main text, this waveplate leads to quantum interference
and Hong-Ou-Mandel photon bunching, not only of two photons but also
of higher photon number states. In particular to also model partially
distinguishable photons, the effects of WP2 has to be modelled carefully
in the computer simulation. We define an $h$-photon Fock state of
$H$-polarized photons in spatial mode $S$ by $|0,h\rangle_{S}$
and an $v$-photon Fock state of $V$-polarized photons in the $L$-mode
by $|v,0\rangle_{L}$. If the photons are completely \emph{distinguishable}
(wave-function overlap $M=0$), the general WP2 transformation is

\begin{equation}
|0,h\rangle_{S}\otimes|v,0\rangle_{L}\xrightarrow[M=0]{WP2(22.5^{\circ})}\frac{1}{\sqrt{v!}\sqrt{h!}}\left(\frac{1}{\sqrt{2}}\right)^{v+h}\sum_{j=0}^{h}\binom{h}{j}(V_{S}{}^{\dagger})^{j}(H_{S}{}^{\dagger})^{h-j}|0\rangle\otimes\sum_{k=0}^{v}\binom{v}{k}(-V_{L}^{\dagger})^{k}(H_{L}^{\dagger})^{v-k}|0\rangle.\label{eq:WP2-dist}
\end{equation}
The photons are individually transformed in 2-dimensional subspaces
of Hilbert space and do not interfere.

On the other hand, quantum interference of completely indistinguishable
photons ($M=1$ and $\left\{ H,V\right\} _{S}=\left\{ H,V\right\} _{L}$)
will lead to photon bunching:

\begin{equation}
|v,h\rangle_{L}\xrightarrow[M=1]{WP2(22.5^{\circ})}\frac{1}{\sqrt{v!}\sqrt{h!}}\left(\frac{1}{\sqrt{2}}\right)^{v+h}\sum_{k=0}^{v}\sum_{j=0}^{h}\binom{v}{k}\binom{h}{j}\left(-1\right)^{k}(V_{L}^{\dagger})^{k+j}(H_{L}^{\dagger})^{h+v-k-j}|0\rangle.\label{eq:WP2-indist}
\end{equation}
In general, WP2 transforms a partially indistinguishable state like
\begin{equation}
\sqrt{M}\cdot|v,h\rangle_{L}+\sqrt{1-M}\cdot|0,h\rangle_{S}\otimes|v,0\rangle_{L}.\label{eq:WP2-dist-indist}
\end{equation}

\subsubsection{Delay loop: Round-trip loss and diffraction\label{subsec:Delay-loop:-Round-trip}}

As usual in quantum optics, we model loss in the delay loop by a beamsplitter,
before WP2, with transmission $t$ and reflection $r$. Ignoring the
empty input port and the dumped output port, this transforms an $n$-photon
input state (single polarization) as
\begin{equation}
|n\rangle\xrightarrow[\text{}]{}\frac{1}{\sqrt{n!}}\sum_{k=0}^{n}(ir)^{n-k}t^{k}(a^{\dagger})^{k}|0\rangle.\label{eq:LoopLoss}
\end{equation}

\begin{figure}[H]
	\begin{centering}
		(a)\includegraphics[width=0.4\columnwidth]{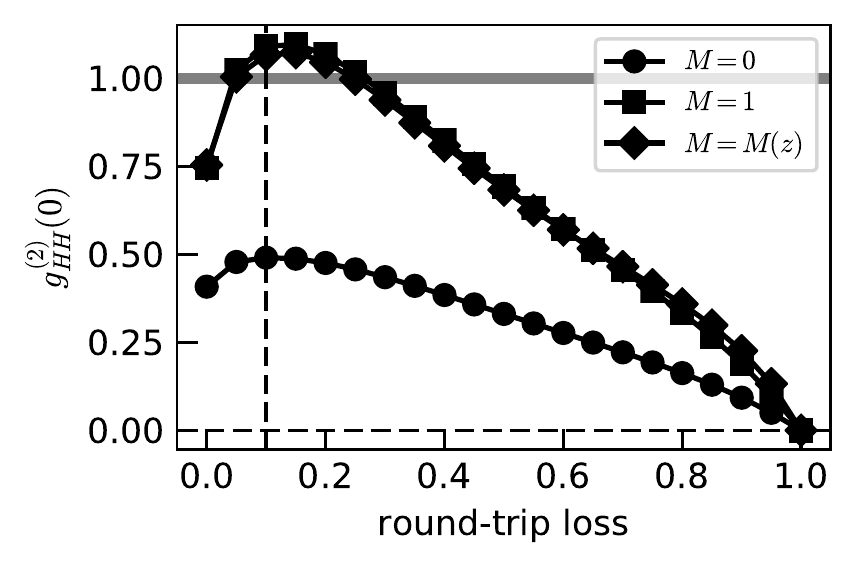}(b)\includegraphics[width=0.4\columnwidth]{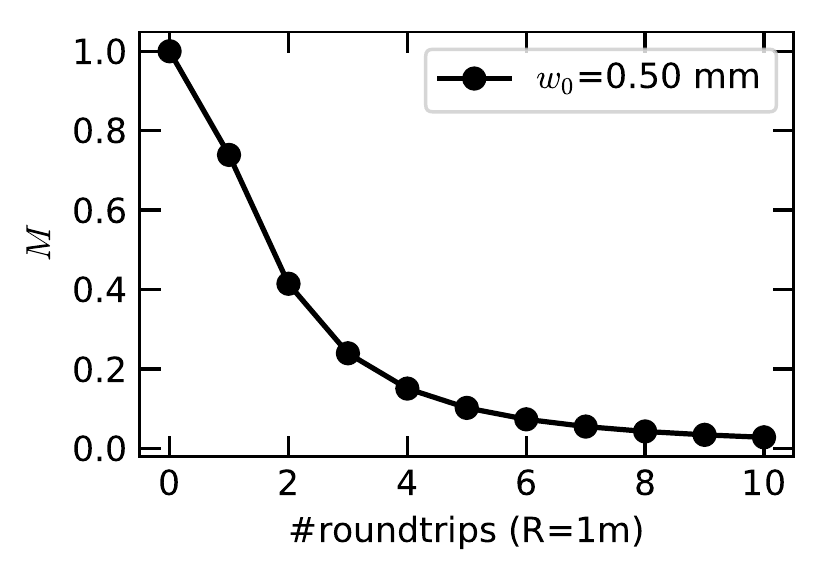}
		\par\end{centering}
	\caption{Effects of round-trip loss and diffraction. (a) Dependence of $g_{HH}^{(2)}(0)$
		on round-trip loss, for the case of fully distinguishable ($M=0$),
		indistinguishable ($M=1$) photons, as well as round-trip dependent
		indistinguishability due to diffraction. The vertical dashed line
		points to our experimental round trip loss of \textasciitilde 0.1.
		(b) Effect of gaussian mode diffraction on the distinguishability
		or wave function overlap $M$. \label{FIG:M-vs-roundtrips}}
\end{figure}

Figure \ref{FIG:M-vs-roundtrips}(a) shows simulation results of $g_{HH}^{(2)}(0)$
for distinguishable and indistinguishable photons, as a function of
round-trip loss. Both curves approach single-photon correlations for
high loss, which is understandable because in this case the delay
loop can be neglected. For low loss, $g_{HH}^{(2)}(0)$ depends strongly
on distinguishability, this is why we can use this as a measure of
quantum interference. 

We do not use relay lenses in the free-space delay loop setup, here
we investigate diffraction between round trips. In order to estimate
the decrease of mode overlap with the number of round trips, we calculate
the propagation-dependent $M$ to be $M(z)=|\frac{kw_{0}^{2}}{kw_{0}^{2}+iz}|^{2}$,
where $k$ is the wave number, $z$ the propagation length, and the
beam waist $w_{0}=0.50\pm0.02\unit{mm}$. This leads to considerably
reduced distinguishability already after 3 round trips as shown in
\ref{FIG:M-vs-roundtrips}(b), however, in combination with the experimental
round-trip loss, its effect on $g_{HH}^{(2)}(0)$ is negligible as
shown in Fig. \ref{FIG:M-vs-roundtrips}(a). 

\subsection{Visibility measurement\label{subsec:Visibility-measurement}}

\begin{figure}[H]
	\begin{centering}
		\includegraphics[width=0.5\columnwidth]{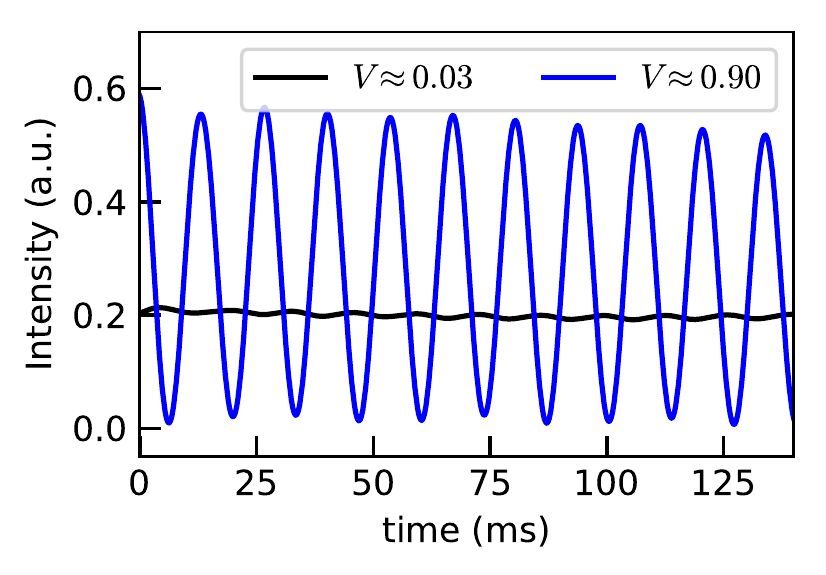}
		\par\end{centering}
	\caption{Background-noise corrected interference measurements for misaligned
		(black) and aligned (blue) delay loop, the obtained visibility $V$
		is indicated in the legend. \label{FIG:Visibility-measurement}}
\end{figure}

We determine the wave-function overlap $M$ on WP2 by measuring the
classical interference visibility $V=\frac{I_{\mathrm{max}}-I_{\mathrm{min}}}{I_{\mathrm{max}}+I_{\mathrm{min}}}$
of laser light sent to the delay loop \cite{Ollivier2020_SM}, where
$I_{\mathrm{max}}$ and $I_{\mathrm{min}}$ are maximal and minimal
intensity. The relative phase is changed simply by scanning the laser
frequency in this unbalanced interferometer. The change in $V$ corresponds
directly to the wave-function overlap $M$ at WP2. We repeat this
measurement before and after the long correlation measurements in
order to determine errors caused by thermal drift of the delay loop
during collecting data, which determines the error bars in Fig. \ref{FIG:HHvsM}
in the main text. Fig. \ref{FIG:Visibility-measurement} shows examples
for visibility measurements with misaligned and aligned delay loop.

\subsection{How many photons do interfere?\label{subsec:How-many-photons}}

\begin{figure}[H]
	\begin{centering}
		(a)\includegraphics[width=0.4\columnwidth]{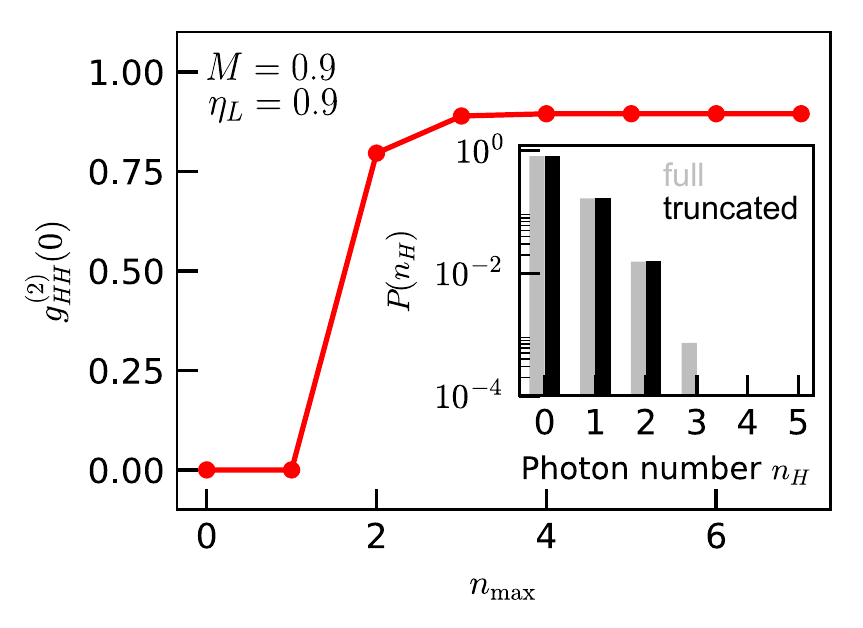}
		(b)\includegraphics[width=0.4\columnwidth]{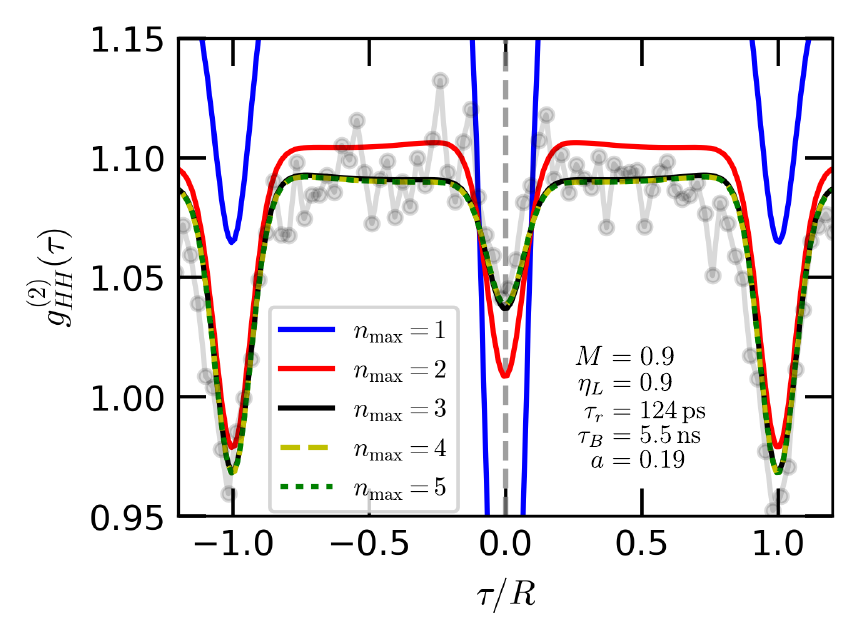}
		\par\end{centering}
	\caption{Influence of higher photon number states based on truncated computer
		simulations. (a) Dependence of $g_{HH}^{(2)}(0)$ on the maximum photon
		number, the inset shows the full (gray) and truncated (black) simulated
		photon number distribution ($M=0.9,\eta_{L}=0.9$). (b) Comparison
		of differently truncated simulations to the experimental $g_{HH}^{(2)}(\tau)$
		data. At least 3 photons are needed to explain the experimental data.
		\label{FIG:Truncated-state-simu}}
\end{figure}

A coherent state contains contributions from a large number of different
photon number states, a natural question about our artificial coherent
states is therefore: What is the highest photon number state that
is required to explain our experimental data? Here we explore this
by truncating out computer simulation and comparing to experimental
data. Figure \ref{FIG:Truncated-state-simu}(a) shows simulated $g_{HH}^{(2)}(0)$
for loop transmission $\eta_{L}=0.9$ and ideal alignment with $M=0.9$.
This state is now truncated to $n_{\mathrm{max}}$ photons and $g_{HH}^{(2)}(0)$
is calculated, see Fig. \ref{FIG:Truncated-state-simu}(a). In Fig.
\ref{FIG:Truncated-state-simu}(b) we show the predicted $g_{HH}^{(2)}(\tau)$
based on Eq. (\ref{eq:G2_HH}). We clearly see that at least 3 photons
are needed to explain our experimental data, but also that discriminating
detection of higher number states is impossible with this method because
of the differences in $g_{HH}^{(2)}(0)$ become negligible. 

This claim is supported by counting simply the number of dips in $g_{HH}^{(2)}(\tau)$
in Fig. \ref{FIG:HH}(b), where clearly dips can be observed at $\tau/R=\pm1$
(corresponding to two photons), $\tau/R=\pm2$ (three photons), and
less clear for four photons. 

Finally, we estimate the rate $R_{n}$ with which $n$-photon states
are produced in our setup. We detect single photons with approximately
$R=150$ kHz, which corresponds to a single-photon rate entering the
HBT setup of approximately $R\cdot2/\eta_{d}$, where $\eta_{d}$
is the single-photon detection efficiency. From our simulation, we
derive the $n$-photon probability $P(n)$, with which we obtain a
three-photon rate of $\approx$ 4.8 kHz, a four-photon rate of $\approx$140
Hz and a 5-photon rate of $\approx1$ Hz.

\subsection{Properties of the artificial coherent state\label{subsec:Artificial-coherent-state-properties}}

\begin{figure}[H]
	\begin{centering}
		\includegraphics[width=0.7\columnwidth]{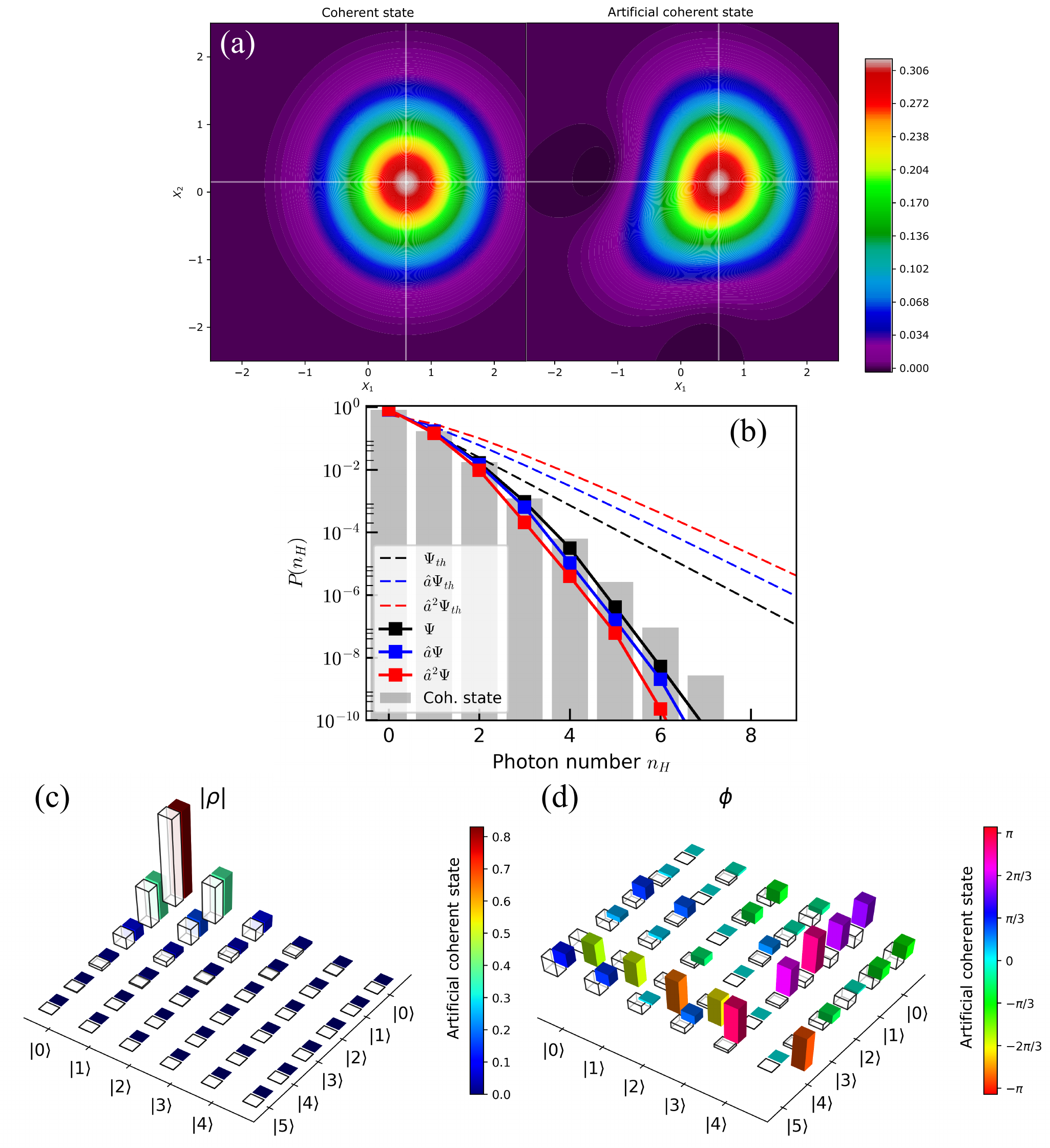}
		\par\end{centering}
	\caption{\textcolor{black}{Comparison of our artificial coherent state to a
			weak coherent state with identical mean photon number ($\bar{n}\approx0.2)$.
			(a) False-color plots of the Wigner function $W(q,p)$ of the artificial
			coherent state (left panel; $M=1,$ round-trip loss 0.1) and of the
			weak coherent state (right panel). Both $W(q,p)$ share the position
			of their maximum (intersection of white lines) as expected. (b) Probability
			distribution after repeated application of the annihilation operator
			$\hat{a}$ on the artificial coherent state ($\Psi$, squares) and
			a weak thermal state ($\Psi_{th}$, dashed lines). Bars show the exact
			coherent state which is an eigenstate of $\hat{a}$.}\textcolor{red}{{}
		}\textcolor{black}{(c) Magnitude and (d) phase of density matrix elements
			of the coherent (wireframe boxes) and the artificial coherent states
			(colored bars). \label{FIG:Wigner}}}
\end{figure}

In the main text, we showed that the artificial coherent state for
$M\approx1$ approaches the photon number distribution of a weak coherent
state and reaches $g^{(2)}(0)\approx1$. In Fig.\ref{FIG:Wigner}(a),
we compare the same states by means of their Wigner function $W(q,p)$,
where $q$ and $p$ are dimensionless conjugate variables corresponding
to electric field quadratures. Both, the position and the value of
the maximum of $W(q,p)$ show that the artificial states are very
close to a coherent state. The presence of negative regions in the
Wigner function evidences nonclassicality, connected to the ability
to create multi-photon entangled states with a delay-loop setup \cite{Istrati2020_SM}. 

Coherent states have the unique property of being eigenstates of the
annihilation operator $\hat{a}$. We test this and show the result
in Fig.\ref{FIG:Wigner}(b), this shows that the artificial coherent
states are very close to being an eigenstate of the annihilation operator,
it is almost unchanged by several applications of $\hat{a}$. On the
other hand, performing the same procedure with a thermal state leads
to pronounced changes, where the vacuum-state probability decreases
strongly and higher number probabilities are increased.
In Fig. \ref{FIG:Wigner}(c) and (d), we compare the density matrices
of the exact weak coherent and artificial coherent states. The similarity
in the density matrix magnitude, Fig. \ref{FIG:Wigner}(c), supports
again the closeness of the states; the difference between both states
appears mostly in the phase and only for the higher photon number
states with low magnitude, see Fig. \ref{FIG:Wigner}(d).

\end{document}